\def\be{\begin{equation}}
\def\ee{\end{equation}}
\def\ba{\begin{array}}
\def\p{\prime}
\def\ea{\end{array}}
\def\t{\tilde}
\def\Rb{{I\!\! R}}
\def\Nb{{I\!\! N}}
\def\Fb{{I\!\! F}}

\documentstyle[12pt]{article}
\begin{document}
\parskip=3pt
\parindent=18pt
\baselineskip=20pt
\setcounter{page}{1}
\centerline{\Large\bf Current Algebraic Structures over Manifolds:}
\vspace{2ex}
\centerline{\Large\bf Poisson Algebras, q-Deformations and Quantization}
\vspace{6ex}
\centerline{\large{\sf Sergio Albeverio$^\star$} ~~~and~~~ {\sf Shao-Ming Fei}
\footnote{\sf Alexander von Humboldt-Stiftung fellow.\\
\hspace{5mm}On leave from Institute of Physics, Chinese Academy of Sciences,
Beijing}}
\vspace{4ex}
\parindent=40pt
{\sf Institute of Mathematics, Ruhr-University Bochum,
D-44780 Bochum, Germany}\par
\parindent=35pt
{\sf $^\star$SFB 237 (Essen-Bochum-D\"usseldorf); 
BiBoS (Bielefeld-Bochum);\par
\parindent=40pt
CERFIM Locarno (Switzerland)}\par
\vspace{6.5ex}
\parindent=18pt
\parskip=5pt
\begin{center}
\begin{minipage}{5in}
\vspace{3ex}
\centerline{\large Abstract}
\vspace{4ex}
Poisson algebraic structures on current
manifolds (of maps from a finite dimensional Riemannian manifold into 
a 2-dimensional manifold) are investigated in terms of symplectic 
geometry. It is shown that there is a one to one correspondence
between such current manifolds and Poisson current algebras 
with three generators. A geometric meaning is given to q-deformations
of current algebras. The geometric quantization of current algebras 
and quantum current algebraic maps is also studied.
\end{minipage}
\end{center}

\newpage

\section{Introduction}
The basic ideas for the subject of current algebras were introduced in
particle physics about 30 years ago\cite{gell}. The
primary ingredients of current algebras are sets of equal-time commutation
relations for the physically conserved currents that arise in the
electromagnetic and weak interactions of hadrons\cite{current}.
Formulae relating $\beta$ decay parameters to pion-nucleon scattering
quantities were physically derived as applications of the current algebra
approach\cite{aw}.

Mathematically, current algebras are maps from a (compact) manifold $N$ to
an algebra $g$. When $N$ is the one dimensional manifold $S^1$, 
current algebras are usually called loop algebras(see e.g.
\cite{abook,eg,pressley,mic}). The representation theory of current algebras
has been studied quite extensively, see e.g. \cite{abook,pressley,mic}

In connection with the theory of integrable systems certain infinite
dimensional ``classical Poisson algebras" have been investigated, see e.g.
\cite{DuN,DukN} and references therein. In the present paper we study
certain current algebras associated with maps from a finite Riemannian 
manifold into a two-dimensional (i.e. 2-D) Riemannian manifold.
We associate to them certain infinite dimensional  Poisson algebras and
characterize them using the symplectic geometry of the underlying
2-D manifold. We then go over to study q-deformations as well as
geometric quantization of these algebras. Before going into more details
about our constructions in infinite dimensions, let us recall how the 
corresponding constructions in finite dimensions are done.
In \cite{fa} it has been shown that there is a one to one correspondence 
between 2-D manifolds and Poisson algebras with three 
generators. This correspondence uses symplectic geometry and gives 
rise to a general description of algebraic structures on 2-D manifolds, 
including Lie algebras and q-deformed Lie algebras with three generators. The
symmetries described by the algebras on 2-D manifolds, the related geometric 
meanings and the geometric meanings of q-deformations are manifest from the 
corresponding 2-D manifolds. The Poisson algebraic maps can be easily 
studied in terms of the corresponding 2-D manifolds. 
For the Lie algebras $SU(2)$ and the quantum (q-deformed)
Lie algebras $SU_q(2)$, the associated quantum algebraic maps between
$SU(2)$ and $SU_q(2)$ are discussed by using the method of 
prequantization and geometric quantization \cite{fg}.

In this paper, we give an infinite dimensional extension of these 
connections. More precisely we study  current algebraic structures 
on 2-D manifolds with parameters taking values in finite dimensional 
Riemannian manifolds. We call these manifolds ``2-D current manifolds".
In the case where the parameter space is $S^1$ we obtain loop algebraic 
structures on 2-D current manifolds. 
We give a complete description of algebraic structures on 2-D 
current manifolds in terms of the underlying symplectic geometry.
We show that there is a one to one correspondence 
between 2-D current manifolds and Poisson current algebras 
with three generators, including current Lie algebras and q-deformed 
current Lie algebras.
The symmetry described by the current algebra and the
geometric meaning of the q-deformation of the current algebra
are manifest from the related current manifolds.
By geometric quantization, we get corresponding quantized current algebras
and related current manifolds.
Both the Poisson current algebraic maps 
and quantum current algebraic maps can then be investigated
in terms of the corresponding classical and quantum current manifolds.

We first recall in section 2 some notations of infinite
dimensional symplectic geometry referring to \cite{marsden} and \cite{wood} 
for background. 
In section 3 we study the symplectic geometry on 2-D current manifolds and
establish relations between Poisson current structures and
2-D current manifolds. 
As applications, we discuss in section 4 some 
special 2-D current manifolds and their related Poisson current manifolds.
Section 5 is dedicated to Poisson current algebraic maps in terms of the 
corresponding current manifolds. We investigate the geometric quantization of
current algebras in section 6 and give some concluding remarks in section 7.

\section{Symplectic Geometry on Current Manifolds}

The basic object in symplectic geometry is a symplectic manifold 
which is an even dimensional manifold equipped with a symplectic two 
form, see e.g. \cite{wood,geom}. Let $M$ be a connected even 
dimensional Riemannian manifold and $N$ an arbitrary 
finite dimensional Riemannian manifold equipped with a finite reference 
measure $\mu$. A current manifold $M_N$ is the space of
smooth mappings from $N$ to $M$, which can be equipped with the topology
of a Banach manifold, see e.g. \cite{pressley}.
Let $\delta$ denote the exterior derivative on $M$. 
By definition a symplectic form $\omega$ on $M_N$, is a two form on $M$ 
with parameters in $N$, which is
\parskip=0pt\par
(i) closed
\be\label{1} 
\delta\omega=0
\ee

(ii) non-degenerate, i.e.,
\be\label{2}
X\rfloor\omega=0\Rightarrow X=0\,,
\ee
where $X$ are vector fields on $M_N$ and $\rfloor$ denotes the left inner
product defined by $(X\rfloor\omega)(Y)=\omega(X,Y)$
for any two smooth vector fields $X$ and $Y$ on $M_N$. It is possible to
show that such symplectic forms exist on $M_N$, see e.g. \cite{marsden} and
\cite{wood} (for the case $dim(M)=2$ we discuss this below).
\parskip=3pt

Canonical transformations are by definition $\omega$-preserving 
diffeomorphisms of $M_N$
onto itself. A vector $X$ on $M_N$ corresponds to an infinitesimal
canonical transformation if and only if the Lie derivative of $\omega$ with
respect to $X$ vanishes,
\begin{equation}\label{3}
{\cal L}_{X}\omega=X\rfloor \delta\,\omega+\delta\,(X\rfloor\omega)=0\,.
\end{equation}
A vector $X$ satisfying (\ref{3}) is said to be a Hamiltonian 
vector field. Since $\omega$ is closed, it follows that a vector $X$ is
a Hamiltonian vector field if and only if $X\rfloor \delta$ is closed.
Let ${\cal F}(M_N)$ denote the real-valued smooth functions on $M_N$. 
For $f\in {\cal F}(M_N)$, since $\omega$ is non-degenerate 
there exists a Hamiltonian vector field $X_{f}$ (unique up to a sign on 
the right hand side of the following equation) satisfying
\begin{equation}\label{4}
X_{f}\rfloor\omega=-\delta\,f\,.
\end{equation}

The Poisson bracket $[f,g]_{P.B.}$ of two smooth functions $f$ and $g$ in
${\cal F}(M_N)$ is defined to be the function $-\omega(X_{f},X_{g})$. It
satisfies the identities:
\begin{equation}\label{5}
[f,g]_{P.B.}=-\omega(X_{f},X_{g})=\omega(X_{g},X_{f})=-X_{f}\,g=X_{g}\,f\,.
\end{equation}

According to Whitney's embedding theorem we can smoothly embed $N$ resp.
$M$ in Euclidean spaces of least possible dimensions $n$ 
($\le 2(dim\,N+1))$ resp. $m$ ($\le 2(dim\,M+1))$, see e.g. \cite{atiyah}.
Let $\vec{x}$ (resp. $S$) be local coordinates of the 
so embedded manifold $N$ (resp. $M$). 
Let $S_i$, $i=1,...,m$, be the components of $S$.
The basis of the tangent vectors (resp. cotangent vectors) of 
the Banach manifold $M_N$ are then
$\{\delta/\delta S_i (\vec{x})\}$ (resp. $\{\delta S_i (\vec{x})\}$),
$i=1,...,m$. For fixed $\vec{x}$, $S_1 (\vec{x}),...,S_m (\vec{x})$
can be looked upon as orthogonal smooth
vectors spanning the tangent space at $\vec{x}\in N$ to $M$. In analogy with 
the finite dimensional situation (see e.g. \cite{wood}) we can define 
an inner product between the bases of tangent vectors on $M_N$ and two forms 
on $M_N$
$$
\frac{\delta}{\delta S_i (\vec{x})}\rfloor 
\delta S_j (\vec{y})\wedge \delta S_k (\vec{z})=
\delta_{ij}{\bf\delta}(\vec{x}-\vec{y}) \delta S_k (\vec{z})
-\delta_{ik}{\bf\delta}(\vec{x}-\vec{z}) \delta S_j (\vec{y})\,,
~~~i,j,k=1,...m\,,
$$
where ${\bf\delta}(\cdot)$ is the usual $\bf\delta$ function on $N$
and $\delta_{ij}=1$ for $i=j$, $\delta_{ij}=0$ for $i\neq j$. This equality is
to be understood in the sense of generalized functions (using the natural
pairing given by the Riemann-Lebesgue volume measure on $N$ and $M$).

\section{Poisson Current Algebraic Structures on 2-D Current Manifolds}

In the following we take $M$ to be a two dimensional Riemannian manifold
smoothly embedded into $\Rb^3$.
Let, as in section 2, $S_i$, $i=1,2,3$, be the coordinates of $M$ 
in $\Rb^3$ and $\vec{x}$ the coordinate vector of the manifold $N$ 
in $\Rb^n$ ($n$ is also as in section 2). 
We consider a general 2-D current manifold $M_N$ defined in terms of some 
smooth real-valued function $F$ on $\Rb^3$ by
\be\label{6}
F(\stackrel{\circ}{S}(\vec{x}))=0\,,~~~~\vec{x}\in N,
\ee
$\stackrel{\circ}{S}$ denoting the values of $S=(S_1,S_2,S_3)$ for which
(\ref{6}) holds. The Poisson algebraic structure
on the current manifold (\ref{6}) is determined by the 
corresponding symplectic structure on it.

Let for general $S$
$$
\Fb(S)\equiv\int_N F(S(\vec{x}))d\mu(\vec{x})\,,
$$
where $\mu$ is the Riemann-Lebesgue volume measure of the Riemannian 
manifold $N$. We look at $\Fb$ as a real-valued smooth functional of 
$S=(S_i,~i=1,2,3)$. When $S$ takes values on $M_N$, $\Fb(S)=0$.
We define $\delta \Fb/\delta S_j(\vec{x})$
as the smooth functional of $S$ s.t. for $h_j\in C(N)$
$$
\int h_j(\vec{x})\frac{\delta \Fb(S)}{\delta S_j(\vec{x})}d\mu(\vec{x})
=\delta \Fb (S;h_j)\equiv\lim_{\epsilon\downarrow 0}
\frac{\Fb(S^{\epsilon h_j})-\Fb(S)}{\epsilon}
$$
with $S^{\epsilon h_j}=(..., S_j+\epsilon h_j, ...)$, $j=1,2,3$.
$\delta \Fb (S;h_j)$ is thus the derivative of $\Fb$ at $S$ in the direction 
$h_j$. We assume that $N$ is compact (or that $\mu$ and the 
functions to be integrated against it which occur in our formulae 
are such that all integrals are finite).
								  
{\sf [Proposition 1].} The Hamiltonian vector fields associated with 
$S_i(\vec{x})$, $i=1,2,3$, on manifold $M_N$ are given by
\be\label{7}
X_{S_i(\vec{x})}=\alpha\sum_{jk=1}^{3}\epsilon_{ijk}
\frac{\delta \Fb(S)}{\delta S_j(\vec{x})}\frac{\delta}{\delta S_k(\vec{x})}\,,
\ee
where $\alpha$ is a real constant and $\epsilon_{ijk}$ is the completely
antisymmetric tensor.

{\sf [Proof].} By definition a Hamiltonian vector field $X_f$ associated
with $f\in{\cal F}(M_N)$ should satisfy equation (\ref{4}).
Let us suppose that the Hamiltonian vector fields 
$X_{S_i(\vec{x})}$ resp. the symplectic form on $M_N$ have the 
following general forms:
\be\label{8}
X_{S_i(\vec{x})}=\sum_{jk=1}^{3}\epsilon_{ijk}
A_j(\vec{x})\frac{\delta}{\delta S_k(\vec{x})}\,,
\ee
resp.
\be\label{9}
\omega=-\frac{1}{2}\sum_{lmn=1}^{3}\int_N\epsilon_{lmn}B_l(\vec{y}) 
\delta S_m(\vec{y})\wedge\delta S_n(\vec{y})d\mu(\vec{y})\,,
\ee
where $A_i(\vec{x})$ and $B_i(\vec{x})$, $i=1,2,3$, are some
smooth functions of $S(\vec{x})$ that will be determined by eq.(\ref{4}). 
Substituting them into equation (\ref{4})
we have (in the sense of generalized functions):
\be\label{10}
\begin{array}{rcl}
X_{S_i(\vec{x})}\rfloor\omega
&=&-\displaystyle\frac{1}{2}
\displaystyle\sum_{jklmn=1}^{3}
\displaystyle\int_N\epsilon_{ijk}\epsilon_{lmn}A_j(\vec{x}) B_l(\vec{y}) 
{\bf \delta}(\vec{x}-\vec{y})[\delta_{km}\delta S_n(\vec{y})
-\delta_{kn}\delta S_m(\vec{y})]d\mu(\vec{y})\\[4mm]
&=&-\displaystyle\frac{1}{2}
\displaystyle\sum_{jlmn=1}^{3}\epsilon_{lmn}A_j(\vec{x}) B_l(\vec{x})
[\epsilon_{ijm}\delta S_n(\vec{x})-\epsilon_{ijn}\delta S_m(\vec{x})]\\[4mm]
&=&-\displaystyle\sum_{jlmn=1}^{3}
\epsilon_{ijm}\epsilon_{lmn}A_j(\vec{x}) B_l(\vec{x})\delta S_n(\vec{x})
\\[5mm]
&=&-\delta S_i(\vec{x})\,,~~~i=1,2,3.
\end{array}
\ee
For $i=1$ we have $X_{S_1(\vec{x})}\rfloor\omega=-\delta S_1(\vec{x})$, i.e.,
$$
-(A_2(\vec{x}) B_2(\vec{x})+A_3(\vec{x}) B_3(\vec{x}))\delta S_1(\vec{x}) 
+A_2(\vec{x}) B_1(\vec{x})\delta S_2(\vec{x})+ 
A_3(\vec{x}) B_1(\vec{x})\delta S_3(\vec{x})
=-\delta S_1(\vec{x})\,.
$$
$\delta S_i(\vec{x})$, $i=1,2,3$, are not linearly independent, in fact
from eq.(\ref{6}) we have
\be\label{11}
\sum_{i=1}^3\left.\frac{\delta \Fb(S)}{\delta S_i}\delta S_i\right\vert_N=0\,.
\ee
Using relation (\ref{11}) we get
$$
A_2(\vec{x})=\alpha\frac{\delta \Fb(S)}{\delta S_2(\vec{x})}\,,~~~
A_3(\vec{x})=\alpha\frac{\delta \Fb(S)}{\delta S_3(\vec{x})}
$$
for some real constant $\alpha$ and
$$
\alpha\left(
B_1(\vec{x})\frac{\delta \Fb(S)}{\delta S_1(\vec{x})}+
B_2(\vec{x})\frac{\delta \Fb(S)}{\delta S_2(\vec{x})}+
B_3(\vec{x})\frac{\delta \Fb(S)}{\delta S_3(\vec{x})}\right)=1\,.
$$
Combining this with the corresponding results from (\ref{10}) with $i=2,3$ 
we obtain
\be\label{13}
A_i(\vec{x})=\alpha\frac{\delta \Fb(S)}{\delta S_i(\vec{x})}\,,~~~i=1,2,3
\ee
and
\be\label{14}
\alpha\sum_{i=1}^3B_i(\vec{x})\frac{\delta \Fb(S)}{\delta
S_i(\vec{x})}=1\,.
\ee
Substituting eq.(\ref{13}) into (\ref{8}) we get (\ref{7}).
\hfill $\rule{2mm}{2mm}$

{\sf [Remark 1].} It is seen that $A_i(\vec{x})$ is independent of the 
coefficients $B_i(\vec{x})$, that is, the Hamiltonian vector field 
associated with
$S_i(\vec{x})$ is independent of the construction of the symplectic form 
$\omega$ on $M_N$. Owing to the equivalence of Hamiltonian vector fields
in 2-D case \cite{fa}, the factor $\alpha$ will be taken to be $1/2$ in the
following.

{\sf [Proposition 2].} The two form $\omega$ given by (\ref{9}) is a 
symplectic form on $M_N$ iff $B_i(\vec{x})$, $i=1,2,3$, satisfy 
condition (\ref{14}).

{\sf [Proof].} It is manifest from the proof above that condition (\ref{14})
is necessary and sufficient for $\omega$ to satisfy formula (\ref{4}).
And due to the fact that $M$ is a 2-D manifold, $\omega$ is obviously 
closed, i.e., $\delta\omega=0$. \hfill $\rule{2mm}{2mm}$ 

On finite dimensional manifolds with a symplectic structure
one can define Poisson algebraic structures.
In a similar way we define the
``Poisson current algebraic structures" on current manifolds equipped 
with a symplectic structure.

{\sf [Theorem 1].} The Poisson current algebraic structure
on the manifold $M_N$ is (uniquely) given by
\be\label{15}
[S_i(\vec{x}),S_j(\vec{y})]_{P.B.}=\frac{1}{2}
\sum_{k=1}^3\epsilon_{ijk}\frac{\delta \Fb(S)}{\delta S_k(\vec{x})}
{\bf\delta}(\vec{x}-\vec{y})\,,
\ee
the equality being in the sense of generalized functions over $N$.

{\sf [Proof].} From formula (\ref{5}) and Proposition 1 we have
$$
[S_i(\vec{x}),S_j(\vec{y})]_{P.B.}=-X_{S_i(\vec{x})}S_j(\vec{y})
=\frac{1}{2}\sum_{k=1}^3\epsilon_{ijk}\frac{\delta \Fb(S)}{\delta S_k(\vec{x})}
{\bf\delta}(\vec{x}-\vec{y}),
$$
where $\alpha$ in formula (\ref{7})
has been taken to be $1/2$. \hfill $\rule{2mm}{2mm}$ 

This Poisson current algebra is independent
of the symplectic form on $M_N$, and is uniquely given by the 
current manifold $\Fb(S)=0$ under the algebraic equivalence \cite{fa}.

{\sf [Proposition 3].} For $f\in {\cal F}(M_N)$, the Hamiltonian vector 
field associated with $f$ is given by
\be\label{16}
X_f=\frac{1}{2}\int_N\sum_{ijk=1}^3\epsilon_{ijk}
\frac{\delta f}{\delta S_i(\vec{y})}\frac{\delta \Fb(S)}{\delta S_j(\vec{y})}
\frac{\delta}{\delta S_k(\vec{y})}d\mu(\vec{y})\,.
\ee

{\sf [Proof].} What we have to prove is that $X_f\rfloor\omega=-\delta f$. 
From
$$
-\delta f=-\int_N\sum_{i=1}^3\frac{\delta f}{\delta S_i(\vec{x})}
\delta S_i(\vec{x})d\mu(\vec{x})\,,
$$
and Proposition 1 we have
$$
-\delta S_i(\vec{x})=X_{S_i(\vec{x})}\rfloor\omega=
\frac{1}{2}
\sum_{jk=1}^3\epsilon_{ijk}\frac{\delta \Fb(S)}{\delta S_j(\vec{x})}
\frac{\delta}{\delta S_k(\vec{x})}\rfloor\omega\,.
$$
Therefore 
$$
-\delta f=\frac{1}{2}\int_N\sum_{ijk=1}^3\epsilon_{ijk}
\frac{\delta f}{\delta S_i(\vec{y})}\frac{\delta \Fb(S)}{\delta S_j(\vec{y})}
\frac{\delta}{\delta S_k(\vec{y})}d\mu(\vec{y})\rfloor\omega\,.
$$
Comparing above formula with the condition $X_f\rfloor\omega=-\delta f$ we get 
formula (\ref{16}). \hfill $\rule{2mm}{2mm}$ 

{\sf [Theorem 2].} For $f,g\in {\cal F}(M_N)$, the Poisson
bracket of $f$ and $g$ is given by
\be\label{17}
[f,g]_{P.B.}=-\frac{1}{2}\int_N\sum_{ijk}^3\epsilon_{ijk}
\frac{\delta f}{\delta S_i(\vec{y})}\frac{\delta \Fb(S)}{\delta S_j(\vec{y})}
\frac{\delta g}{\delta S_k(\vec{y})}d\mu(\vec{y})\,.
\ee

{\sf [Proof].} This is a direct result of Proposition 3 and formula (\ref{5}).
\hfill $\rule{2mm}{2mm}$ 

Generally, Poisson current algebras are by definition maps from the compact 
manifold $N$ to finite dimensional algebras. 
Let us consider a general Poisson current algebra
\be\label{b1}
[S_i(\vec{x}),S_j(\vec{y})]_{P.B.}=
\sum_{k=1}^3\epsilon_{ijk}f_k(S(\vec{x})){\bf\delta}(\vec{x}-\vec{y})\,,
~~~\vec{x},\vec{y}\in N,~~~S\in\Rb^3
\ee
(in the sense of distribution on $N$),
where $f_i$, $i=1,2,3$, are smooth functions of $S$ taking values on a
finite dimensional manifold $N$ and $S_i(\vec{x})$, $i=1,2,3$,
satisfy the Jacobi identity.

{\sf [Definition].} If the $f_i$, $i=1,2,3$, satisfy
\be\label{b2}
\left.\frac{\partial f_i}{\partial S_j}\right\vert_p=
\left.\frac{\partial f_j}{\partial S_i}\right\vert_p\,,
~~~~p\in N,~~~ i,j=1,2,3,
\ee
then the algebra (\ref{b1}) is said to be integrable.

{\sf [Remark 2].} The integrable condition (\ref{b2}) is a sufficient
condition for a general Poisson current algebra (\ref{b1}) to satisfy the
Jacobi identity,
$$
\ba{l}
[S_1(\vec{x}),[S_2(\vec{y}),S_3(\vec{z})]_{P.B.}]_{P.B.}
+[S_2(\vec{y}),[S_3(\vec{z}),S_1(\vec{x})]_{P.B.}]_{P.B.}
+[S_3(\vec{z}),[S_1(\vec{x}),S_2(\vec{y})]_{P.B.}]_{P.B.}\\[4mm]
=\left.\left[\displaystyle\frac{\partial f_1}{\partial S_2}f_3
-\displaystyle\frac{\partial f_1}{\partial S_3}f_2+
\displaystyle\frac{\partial f_2}{\partial S_3}f_1
-\displaystyle\frac{\partial f_2}{\partial S_1}f_3+
\displaystyle\frac{\partial f_3}{\partial S_1}f_2
-\displaystyle\frac{\partial f_3}{\partial S_2}f_1\right]
\right\vert_{\vec{x}}
{\bf\delta}(\vec{x}-\vec{y}){\bf\delta}(\vec{y}-\vec{z})=0\,.
\ea
$$

{\sf [Remark 3].} 
Comparing formula (\ref{15}) with formula (\ref{b1}) we see that the
$f_i(S(\vec{x}))$ in the Poisson current bracket (\ref{15}) on $M_N$ is
given by
\be\label{fi}
f_i(S(\vec{x}))=\frac{1}{2}\frac{\delta \Fb(S)}{\delta S_i(\vec{x})},
\ee
where $\Fb(S)=\int_N F(S(\vec{x}))d\mu(\vec{x})$.
As $F\in{\cal F}$ we have
$$
\left.\frac{\partial^2 F(S)}{\partial S_i\partial S_j}\right\vert_p=
\left.\frac{\partial^2 F(S)}{\partial S_j\partial S_i}\right\vert_p
~~~~p\in N,~~~ i,j=1,2,3.
$$
Therefore $f_i$ in (\ref{fi}) satisfy the integrable condition (\ref{b2})
and all the Poisson current algebras 
(\ref{15}) in Theorem 1, over 2-D current manifolds are integrable ones.

{\sf [Theorem 3].} Let $M,N$ be Riemannian manifolds smoothly embedded
in $\Rb^3$ resp. $\Rb^n$. For a given integrable current
Poisson algebra (\ref{b1}),
there exists a 2-D symplectic current manifold $M_N$ 
described by an equation of the form 
$\int_{N}F(S(\vec{y}))d\vec{y}=c$, with $S\in M$, $\vec{y}\in N$, 
$F\in {\cal F}$ and $c$ an arbitrary real number, such that the
Poisson current algebra generated by $S(\vec{y})$ coincides with the
algebra (\ref{b1}).

{\sf [Proof].} A general integrable Poisson algebra is of the form
(\ref{b1}),
$$
[S_i(\vec{x}),S_j(\vec{y})]_{P.B.}=
\sum_{k=1}^3\epsilon_{ijk}f_k(S(\vec{x})){\bf\delta}(\vec{x}-\vec{y})\,,
~~~\vec{x},\vec{y}\in N,~~~S\in\Rb^3,
$$
where $f_i$, $i=1,2,3$, satisfy the integrability condition (\ref{b2}).
What we have to show is that this Poisson algebra can be described by
the symplectic geometry on a suitable 2-D symplectic current
manifold ($M_N,\omega$), in
the sense that the above Poisson current bracket can be described by the
formula (\ref{5}), i.e., the Poisson current bracket 
$[S_i(\vec{x}),S_j(\vec{y})]_{P.B.}$ is given by the Hamiltonian
vector field $X_{S_i(\vec{x})}$ associated with $S_i(\vec{x})$ such that
\be\label{th1}
[S_i(\vec{x}),S_j(\vec{y})]_{P.B.}=-X_{S_i(\vec{x})}S_j(\vec{y})=
\sum_{k=1}^3\epsilon_{ijk}f_k(S(\vec{x})){\bf\delta}(\vec{x}-\vec{y})\,.
\ee

Let $X_{S_i(\vec{x})}^\p$ be given by
\be\label{a3}
X_{S_i(\vec{x})}^\p\equiv\sum_{jk=1}^{3}\epsilon_{ijk}
f_j(\vec{x})\frac{\delta}{\delta S_k(\vec{x})}\,.
\ee

Then $X_{S_i(\vec{x})}^\p$ satisfies
$$
[S_i(\vec{x}),S_j(\vec{y})]_{P.B.}=-X_{S_i(\vec{x})}^\p S_j(\vec{y})=
\sum_{k=1}^3\epsilon_{ijk}f_k(S(\vec{x})){\bf\delta}(\vec{x}-\vec{y})\,.
$$

A general two form on $\Rb^3_N$ has the form
\be\label{wp}
\omega^\p=-\frac{1}{2}\sum_{lmn=1}^{3}\int_N\epsilon_{lmn}B_l(\vec{y}) 
\delta S_m(\vec{y})\wedge\delta S_n(\vec{y})d\mu(\vec{y})\,,
\ee
where $B_i(\vec{x})$, $i=1,2,3$, are some
smooth functions of $S(\vec{x})$.

We have to prove that $S$ can be restricted to a suitable 2-dimensional 
manifold $M\subset\Rb^3$ in such a way that $X_{S_i(\vec{x})}^\p$ coincides 
with the Hamiltonian vector field $X_{S_i(\vec{x})}$ and $\omega^\p$ is the
corresponding symplectic form $\omega$ on $M_N$.

A two form on a 2-D current
manifold is always closed. What we should then check is that
there exists $M\subset\Rb^3$ such that for $S$ restricted to $M$
the formula (\ref{4}) holds for $f=S_i(\vec{x})$, i.e.,
\be\label{a2}
X_{S_i(\vec{x})}^\p\rfloor\omega^\p=-\delta S_i(\vec{x}),~~~i=1,2,3.
\ee

Substituting formulae (\ref{a3}) and (\ref{wp}) into (\ref{a2}) we get
$$
\begin{array}{rcl}
X_{S_i(\vec{x})}^\p\rfloor\omega^\p
&=&-\displaystyle\frac{1}{2}
\displaystyle\sum_{jlmn=1}^{3}\epsilon_{lmn}f_j(\vec{x}) B_l(\vec{x})
[\epsilon_{ijm}\delta S_n(\vec{x})-\epsilon_{ijn}\delta S_m(\vec{x})]\\[4mm]
&=&-\displaystyle\sum_{jlmn=1}^{3}
\epsilon_{ijm}\epsilon_{lmn}f_j(\vec{x}) B_l(\vec{x})\delta S_n(\vec{x})\\[5mm]
&=&-\delta S_i(\vec{x})\,.
\ea
$$

That is,
\be\label{a4}
\ba{l}
(1-f_2(\vec{x})B_2(\vec{x})-f_3(\vec{x})B_3(\vec{x}))
\delta S_1(\vec{x})+f_2(\vec{x})B_1(\vec{x})\delta S_2(\vec{x})+
f_3(\vec{x})B_1(\vec{x})\delta S_3(\vec{x})=0,\\[3mm]
(1-f_3(\vec{x})B_3(\vec{x})-f_1(\vec{x})B_1(\vec{x}))
\delta S_2(\vec{x})+f_3(\vec{x})B_2(\vec{x})\delta S_3(\vec{x})+
f_1(\vec{x})B_2(\vec{x})\delta S_1(\vec{x})=0,\\[3mm]
(1-f_1(\vec{x})B_1(\vec{x})-f_2(\vec{x})B_2(\vec{x}))
\delta S_3(\vec{x})+f_1(\vec{x})B_3(\vec{x})\delta S_1(\vec{x})+
f_2(\vec{x})B_3(\vec{x})\delta S_2(\vec{x})=0.
\ea
\ee

Let us now look at the coefficient determinant $D$ of the 
$\delta S_i(\vec{x})$ in
the system (\ref{a4}). By a suitable choice of $B_1,B_2,B_3$ we can
obtain that $D$ is zero. This is in fact equivalent with the equation
\be\label{a5}
f_1(\vec{x})B_1(\vec{x})+f_2(\vec{x})B_2(\vec{x})
+f_3(\vec{x})B_3(\vec{x})=1
\ee
being satisfied. The fact that $D=0$ implies that there exists indeed an
$M$ as above. 

Substituting condition (\ref{a5}) into (\ref{a4}) we get
\be\label{a6}
f_1(\vec{x})\delta S_1(\vec{x})+f_2(\vec{x})\delta S_2(\vec{x})
+f_3(\vec{x})\delta S_3(\vec{x})=0\,.
\ee

From the assumption (\ref{b2}) we know that the differential
equation (\ref{a6}) is exactly solvable, in the sense that
there exists a smooth (potential) function $F\in {\cal F}(M_N)$ and a
constant $c$ such that
\be\label{a7}
\Fb(S)\equiv\int_{N}F(S(\vec{y}))d\vec{y}=c
\ee
and $\delta\Fb/\delta S_i(\vec{x})=f_i(S(\vec{x}))$. 
The above manifold $M_N$ is then described by (\ref{a7}).

Therefore for any given integrable Poisson current algebra there 
always exists a 2-D current manifold of the form (\ref{a7}) on which
$X_{S_i(\vec{x})}^\p$ in (\ref{a3}) is a Hamiltonian vector field and 
the Poisson bracket of the current algebra is given by 
$X_{S_i(\vec{x})}^\p$ according to the formula,
$$
[S_i(\vec{x}),S_j(\vec{y})]_{P.B.}=-X_{S_i(\vec{x})}^\p S_j(\vec{y})=
\sum_{k=1}^3\epsilon_{ijk}f_k(S(\vec{x})){\bf\delta}(\vec{x}-\vec{y})\,.
$$

The 2-D current manifold defined by (\ref{a7}) is unique (once $c$ is
given). Hence an integrable Poisson current algebra is uniquely given 
by the 2-D current manifold $M_N$ described by 
$\Fb(S)\equiv\int_{N}F(S(\vec{y}))d\vec{y}=c$. $\rule{2mm}{2mm}$

\section{Poisson Algebraic Structures on Some Special Current Manifolds}

In this section we discuss Poisson algebraic structures on some
special current manifolds, which give rise to special current extensions of 
Poisson Lie algebras and q-deformed Lie algebras. In all examples below
$\vec{x}$ takes values in a Riemannian manifold $N$, smoothly embedded in
$\Rb^n$.

a). We first consider a ``current 2-D sphere" given by

\be\label{18}
S_1^2(\vec{x})+S_2^2(\vec{x})+S_3^2(\vec{x})=S_0^2\,,
\ee
where $S_0$ is a real constant$\ne 0$.

$M$ is then here the sphere in $\Rb^3$ of radius $S_0$.
From Theorem 1 we have the Poisson relations (in the sense of
distributions),
\be\label{19}
[S_i(\vec{x}),S_j(\vec{y})]_{P.B.}=\sum_{k=1}^3 \epsilon_{ijk}
S_k(\vec{x})\delta(\vec{x}-\vec{y})\,,
\ee
The algebra defined by (\ref{19}) is a current extension 
of the Poisson algebra $SU(2)$ (cfr. e.g. \cite{gell,current,mic,string}).

A symplectic form on the current manifold $M_N$ 
can be constructed by using formula (\ref{9}) and
condition (\ref{14}). From (\ref{14}) and (\ref{18}) we have
$$
B_1(\vec{x}) S_1(\vec{x})+B_2(\vec{x}) S_2(\vec{x})
+B_3(\vec{x}) S_3(\vec{x})=1\,.
$$
Comparing this with eq.(\ref{18}) we can simply take 
$B_i(\vec{x})=S_i(\vec{x})/S_0^2$.
Therefore from (\ref{9}) we obtain the following symplectic form
\begin{equation}\label{20}
\omega=\frac{-1}{2S_{0}^{2}}\sum_{ijk=1}^3\int_N
\epsilon_{ijk}S_i(\vec{x})\delta S_j(\vec{x})\wedge\delta S_k(\vec{x})
d\mu(\vec{x})
\end{equation}
 
b). We now consider a ``q-deformed current 2-D sphere" defined by
\begin{equation}\label{21}
S_1^2(\vec{x})+S_2^2(\vec{x})
+\frac{(\sinh\gamma S_3(\vec{x}))^{2}}{\gamma \sinh\gamma}
=\frac{(\sinh\gamma S_0)^{2}}{\gamma \sinh\gamma}
\stackrel{def}{=}S_\gamma^2\,,  
\end{equation}
where $\gamma=\log q$, $q\in\Rb\backslash\{0\}$ is the deformation parameter. 
Heuristically, when $\gamma\to 0$, the manifold (\ref{21}) becomes the 
current 2-D sphere (\ref{18}).

For the manifold (\ref{21}), Theorem 1 gives rise to the following Poisson
algebraic relations (again written in the sense of generalized functions):
$$
\begin{array}{rcl}
[S_1(\vec{x}),S_2(\vec{y})]_{P.B.}
&=&\displaystyle\frac{\sinh 2 \gamma S_3(\vec{x})}
{2\sinh\gamma}\delta(\vec{x}-\vec{y})\,,\\[4mm]
[S_2(\vec{x}),S_3(\vec{y})]_{P.B.}&=&S_1(\vec{x})
\delta(\vec{x}-\vec{y})\,,\\[4mm]
[S_3(\vec{x}),S_1(\vec{y})]_{P.B.}&=&S_2(\vec{x})\delta(\vec{x}-\vec{y})\,,
\end{array}
$$
or, equivalently,
\be\label{22} 
\begin{array}{l}
[S_+(\vec{x}),S_-(\vec{y})]_{P.B.}
=-i \displaystyle\frac{\sinh 2\gamma S_3(\vec{x})}
{\sinh\gamma}\delta(\vec{x}-\vec{y})
\,,\\[4mm]
[S_3(\vec{x}),S_{\pm}(\vec{y})]_{P.B.}=\mp i S_{\pm}(\vec{x})
\delta(\vec{x}-\vec{y})\,,
\end{array}
\ee
where $S_{\pm}(\vec{x})=S_{1}(\vec{x})\pm i S_{2}(\vec{x})$ and $i=\sqrt{-1}$.

The algebra (\ref{22}) is just the current extension of the 
q-deformed Poisson Lie algebra $SU_q(2)$\cite{jimbo}.
It is isomorphic (up to a factor i) to the current extension of the 
``quantum" algebra $SU_q(2)$, but is here classically realized, which means 
that the q-deformation and physical
$\hbar$-quantization of the current extended algebras are independent, 
like in the case of Lie algebras\cite{fg,flato}. Both current extended 
Lie algebras and current extended q-deformed Lie algebras can thus be 
realized at classical and also at quantum levels (see section 6).

Condition (\ref{14}) now turns out to be
$$
B_1(\vec{x})S_1(\vec{x})+B_2(\vec{x})S_2(\vec{x})
+B_3(\vec{x})\frac{\sinh 2\gamma S_3(\vec{x})}{2\sinh\gamma}=1\,.
$$
From eq.(\ref{21}) we have
$$
B_1(\vec{x})=\frac{\sqrt{\gamma\sinh\gamma}}{\sinh(\gamma S_0)} S_1(\vec{x})\,,~~~~
B_2(\vec{x})=\frac{\sqrt{\gamma\sinh\gamma}}{\sinh(\gamma S_0)} S_2(\vec{x})\,,~~~~
B_3(\vec{x})=\frac{\sqrt{\gamma\sinh\gamma}}{\gamma\sinh(\gamma S_0)} 
\tanh\gamma S_3(\vec{x})\,.
$$
Substituting these expressions into eq.(\ref{9}) we obtain the symplectic 
form
\begin{equation}\label{23}
\begin{array}{rcl}
\omega&=&
-\displaystyle\frac{\gamma\sinh\gamma}{(\sinh\gamma S_0)^{2}}
\displaystyle\int_N
\left[S_1(\vec{x})\delta S_2(\vec{x})\wedge \delta S_3(\vec{x}) 
+S_2(\vec{x}) \delta S_3(\vec{x}) \wedge \delta S_1(\vec{x})\right.\\[4mm]
&&+\left.\displaystyle\frac{\tanh\gamma S_3(\vec{x})}{\gamma} 
\delta S_1(\vec{x})\wedge \delta S_2(\vec{x})\right]d\mu(\vec{x})\,.
\end{array}
\end{equation}

c). The ``current elliptic paraboloid" is defined by

\be\label{24}
S_1^2(\vec{x})+S_2^2(\vec{x})-S_3(\vec{x})=\frac{1}{2}\,.
\ee
From formula (\ref{15}) we have, in the sense of generalized functions:
\begin{equation}\label{25}
\begin{array}{l}
[S_1(\vec{x}),S_2(\vec{y})]_{P.B.}=-\frac{1}{2}\delta(\vec{x}-\vec{y})\,,\\[4mm]
[S_2(\vec{x}),S_3(\vec{y})]_{P.B.}=S_1(\vec{x})\delta(\vec{x}-\vec{y})\,,\\[4mm]
[S_3(\vec{x}),S_1(\vec{y})]_{P.B.}=S_2(\vec{x})\delta(\vec{x}-\vec{y})\,.
\end{array}
\end{equation}
This is just the current extension of the Poisson simple harmonic oscillator 
algebra ${\cal H}(4)$ \cite{h}.

The corresponding symplectic form can be obtained by using formulae 
(\ref{9}), (\ref{14}) and eq.(\ref{24}),
\be\label{26}
\begin{array}{rcl}
\omega&=&-2\displaystyle\int_N
\left[S_1(\vec{x})\delta S_2(\vec{x})\wedge\delta S_3(\vec{x})
+S_2(\vec{x})\delta S_3(\vec{x})\wedge\delta S_1(\vec{x})\right.\\[4mm]
&&+\left.2S_3(\vec{x})\delta S_1(\vec{x})\wedge\delta S_2(\vec{x})\right]
d\mu(\vec{x})\,.
\end{array}
\ee

d). The ``q-deformed current elliptic paraboloid" is defined by

\begin{equation}\label{27}
S_1^{2}(\vec{x})+S_2^{2}(\vec{x})-\frac{\sinh(2\gamma S_3(\vec{x}))}
{2\gamma\cosh\gamma}
=\frac{\sinh\gamma}{2\gamma\cosh (\gamma)}\,,
\end{equation}
where again $\gamma=\log q$, $q\in\Rb\backslash\{0\}$ is 
the deformation parameter. 

The algebra on this current manifold is
the q-deformed current extension of the simple harmonic oscillator algebra of
${\cal H}_{q}(4)$ considered in \cite{hq} and is described (in the sense
of generalized functions) by:
\begin{equation}\label{28}
\begin{array}{l}
[S_1(\vec{x}),S_2(\vec{y})]_{P.B.}=-\displaystyle
\frac{\cosh (2\gamma S_3(\vec{x}))}
{2\cosh\gamma}\delta(\vec{x}-\vec{y})\,,\\[4mm]
[S_2(\vec{x}),S_3(\vec{y})]_{P.B.}=S_1(\vec{x})\delta(\vec{x}-\vec{y})\,,
\\[4mm]
[S_3(\vec{x}),S_1(\vec{y})]_{P.B.}=S_2(\vec{x})\delta(\vec{x}-\vec{y})\,.
\end{array}
\end{equation}

Similarly as above we have the symplectic form
\be\label{29}
\begin{array}{rcl}
\omega&=&-2\displaystyle\frac{\gamma\cosh\gamma}{\sinh\gamma}
\displaystyle\int_N\left[
S_1(\vec{x})\delta S_2(\vec{x})\wedge\delta S_3(\vec{x})
+S_2(\vec{x})\delta S_3(\vec{x})\wedge\delta S_1(\vec{x})\right.\\[4mm]
&&\left.+\displaystyle\frac{\sinh(2\gamma S_3(\vec{x}))}
{\gamma\cosh(2\gamma S_3(\vec{x}))}\delta S_1(\vec{x})
\wedge\delta S_2(\vec{x})\right]d\mu(\vec{x})\,.
\end{array}
\ee

e). The ``current manifold of the one sheet hyperboloid" is defined by

\begin{equation}\label{30}
S_1^2(\vec{x})+S_2^2(\vec{x})-S_3^2(\vec{x})=constant\,.
\end{equation}
The corresponding current algebra is described (in the sense of generalized
functions) by the relations:
\begin{equation}\label{31}
\begin{array}{l}
[S_1(\vec{x}),S_2(\vec{y})]_{P.B.}=-S_3(\vec{x})\delta(\vec{x}-\vec{y})\,,
\\[4mm]
[S_2(\vec{x}),S_3(\vec{y})]_{P.B.}=S_1(\vec{x})\delta(\vec{x}-\vec{y})\,,
\\[4mm]
[S_3(\vec{x}),S_1(\vec{y})]_{P.B.}=S_2(\vec{x})\delta(\vec{x}-\vec{y})\,.
\end{array}
\end{equation}
This is the current extension of the well known Poisson $SU(1,1)$ algebra. 
The corresponding symplectic form $\omega$ can easily be obtained. In fact 
it is given by (\ref{20}), with $S_0$ being replaced by
the constant on the right hand side of eq.(\ref{30}). This also shows that the
symplectic structures are in general not unique for a given symplectic 
manifold, since there could be more than one set of $B_i(\vec{x})$, $i=1,2,3$ 
satisfying eq.(\ref{14}).

Using formulas (\ref{15}), (\ref{9}) and (\ref{14}),
other examples of current extended Poisson algebras and 
related symplectic structures associated with current manifolds can
be constructed in a similar way, though formally
the current algebras could in general not be the current extensions
of Lie algebras or q-deformed Lie algebras. For example
for the ``genus one current Riemannian manifold",
\be\label{32}
(\sqrt{S_1^2(\vec{x})+S_2^2(\vec{x})}-a)^{2}+S_3^2(\vec{x})=r^{2}\,,
\end{equation}
where $a$ and $r$ are the radii of the torus, we have the following 
Poisson current algebra
\begin{equation}\label{33}
\begin{array}{l}
[S_1(\vec{x}),S_2(\vec{y})]_{P.B.}=S_3(\vec{x})\delta(\vec{x}-\vec{y})\,,
\\[4mm]
[S_2(\vec{x}),S_3(\vec{y})]_{P.B.}=\left[S_1(\vec{x})
-\displaystyle
\frac{aS_1(\vec{x})}{\sqrt{S_1^2(\vec{x})+S_2^2(\vec{x})}}\right]
\delta(\vec{x}-\vec{y})\,,\\[4mm]
[S_3(\vec{x}),S_1(\vec{y})]_{P.B.}=\left[S_2(\vec{x})
-\displaystyle\frac{aS_2(\vec{x})}{\sqrt{S_1^2(\vec{x})+S_2^2(\vec{x})}}
\right]\delta(\vec{x}-\vec{y})
\end{array}
\end{equation}
(in the sense of generalized functions).

\section{Poisson Current Algebraic Maps}

From the above we see that the current algebras which are defined as
maps from a compact manifold $N$ to
an algebra with three generators are related, via symplectic 
geometry, to certain 2-D current manifolds and vice versa. 
Therefore it is convenient to investigate current algebraic 
maps by using the associated 2-D current manifolds.

Let $A_0$ and $A_0^\prime$ be algebras with three generators. Let $N$ and 
$N^\prime$ be Riemannian manifolds smoothly embedded in $\Rb^3$
such that $D\equiv N\cap N^\prime\neq\emptyset$.
Let $A$ and $A^\prime$ be current algebras of 
mappings from $N$ and $N^\prime$ to $A_0$ and $A_0^\prime$ respectively.
Let $S\equiv S_i(\vec{x})$ (resp. 
$S^{\prime}\equiv S_i^\prime (\vec{x})$), $i=1,2,3$, 
$\vec{x}\in D$ be the generators of the current algebra $A$
(resp. $A^\prime$), with corresponding current manifolds 
$F(S)=0$ (resp. $F^\prime(S^\prime)=0$)
defined by a certain smooth real-valued function $F$ (resp. $F^\prime$).

{\sf [Theorem 3].} $\t{S}(S)\equiv\{\t{S}_i(\vec{x}),~i=1,2,3;~\vec{x}\in D\}$ 
generates $A^{\prime}$ iff $\t{S}$ satisfies $\Fb^\prime(\t{S})=0$,
where $\Fb^\prime(\t{S})\equiv\int_D F^\prime(\t{S})d\mu$ and
$\Fb(S)\equiv\int_D F(S)d\mu=0$.

{\sf [Proof].} Let $M_D$ be the manifold defined by the equation
$F(S(\vec{x}))=0$, $\vec{x}\in D$ and $M_D^\prime$ the manifold defined by the equation
$F^\prime (S^\prime(\vec{x}))=0$, $\vec{x}\in D$.
If $\t{S}$ satisfies $\Fb^\prime(\t{S})=0$, then from Theorem 1 $\t{S}$ 
gives rise to the current algebra $A^\prime$.

Conversely we have to prove that if $\t{S}(S)$ generates the current
algebra $A^\prime$ by using the algebraic relations of $S$, then 
$\t{S}$ satisfies the equation $\Fb^\prime(\t{S})=0$.

Due to the relation $F(S)=0$, the $S_i$,
$i=,1,2,3$, are not independent. Since $F(S)=0$ is assumed to be a
two dimensional manifold, we can take, without loosing generality, 
$S_1$, $S_2$ to be the independent variables. 
For $f,g\in {\cal F}(M_D)$, Theorem 2 says that the Poisson bracket of $f$ 
and $g$ is given by
\be\label{a1}
[f,g]_{P.B.}=
\frac{1}{2}\int_D \frac{\delta\Fb(S)}{\delta S_3(\vec{y})}
\left(\frac{\delta f}{\delta S_1(\vec{y})}
\frac{\delta g}{\delta S_2(\vec{y})}-
\frac{\delta f}{\delta S_2(\vec{y})}
\frac{\delta g}{\delta S_1(\vec{y})}\right)
d\mu(\vec{y})\,.
\ee

From Theorem 1 the Poisson current algebra $A^\prime$ is given by
\be\label{e1}
[\t{S}_i(\vec{x}),\t{S}_j(\vec{y})]_{P.B.}=\frac{1}{2}
\sum_{k=1}^3\epsilon_{ijk}\frac{\delta \Fb^\prime(\t{S})}{\delta \t{S}_k(\vec{x})}
{\bf\delta}(\vec{x}-\vec{y})
\ee
(in the sense of generalized functions). On the other hand
from (\ref{a1}), the
Poisson algebraic relations of $\t{S}_i(\vec{x})$ are
\be\label{e2}
[\t{S}_i(\vec{x}),\t{S}_j(\vec{y})]_{P.B.}=\frac{1}{2}
\int_D \frac{\delta\Fb(S)}{\delta S_3(\vec{z})}
\left(\frac{\delta \t{S}_i(\vec{x})}{\delta S_1(\vec{z})}
\frac{\delta \t{S}_j(\vec{y})}{\delta S_2(\vec{z})}-
\frac{\delta \t{S}_i(\vec{x})}{\delta S_2(\vec{z})}
\frac{\delta \t{S}_j(\vec{y})}{\delta S_1(\vec{z})}\right)
d\mu(\vec{z})\,.
\ee
Taking into account that (as generalized functions)
$$ 
\frac{\delta \t{S}_i (\vec{x})}{\delta S_j (\vec{y})}=
\t{S}_{i,j}(\vec{x})
{\bf\delta}(\vec{x}-\vec{y})\,,
$$
(where $\t{S}_{i,j}(\vec{x})$ is the function of $\vec{x}$ obtained by
evaluating the derivative of the function $\t{S}_i$ of $S_j$ with
respect to $S_j$ at $\vec{x}$), we get from (\ref{e1}) and (\ref{e2}):
$$
\frac{\delta\Fb(S)}{\delta S_3(\vec{y})}
\left.\left(\frac{\delta \t{S}_i}{\delta S_1}
\frac{\delta \t{S}_j}{\delta S_2}-
\frac{\delta \t{S}_i}{\delta S_2}
\frac{\delta \t{S}_j}{\delta S_1}\right)\right\vert_{\vec{y}}
{\bf\delta}(\vec{x}-\vec{y})
=\sum_{k=1}^3\epsilon_{ijk}\frac{\delta \Fb^\prime(\t{S})}{\delta \t{S}_k(\vec{x})}
{\bf\delta}(\vec{x}-\vec{y})\,,
$$
By integrating with respect to $d\mu(\vec{y})$ above equations for $i=1,\,j=2$ resp.
$i=2,\,j=3$ resp. $i=3,\,j=1$ and
multiplying the so obtained equations by 
$\frac{\delta S_3^\prime(\vec{x})}{\delta S_k(\vec{z})}$ resp.
$\frac{\delta S_1^\prime(\vec{x})}{\delta S_k(\vec{z})}$ resp.
$\frac{\delta S_2^\prime(\vec{x})}{\delta S_k(\vec{z})}$, 
summing then these equations together and finally integrating with 
respect to $d\mu(\vec{x})$, we get
$$
\frac{\delta \Fb^\prime(\t{S})}{\delta S_k(\vec{z})}=0\,,~~~~~k=1,2, ~~~\forall
\vec{z}\in D.
$$
Therefore $\Fb^\prime(\t{S})=constant$. This is equivalent to the current 
manifold $\Fb^\prime(\t{S})=0$ for the 
algebra $A^\prime$, since a constant term does not change the current 
algebraic structures of the manifold. \hfill  $\rule{2mm}{2mm}$

We give two examples of Poisson current algebraic maps. 
Eqs.(\ref{18}) and (\ref{21}) give algebraic maps 
between the current extended algebras of $SU(2)$ and $SU_{q}(2)$,
\begin{equation}\label{34}
S_{\pm}^{\prime}(\vec{x})=\frac{1}{\sqrt{\gamma\sinh\gamma}}S_{\pm}(\vec{x})
\frac{\sinh\gamma(S_{0}\mp S_{3}(\vec{x}))}{S_{0}
\mp S_{3}(\vec{x})}~,~~~S_{3}^{\prime}(\vec{x})=S_{3}(\vec{x})~,
\end{equation}
where $S_{\pm}(\vec{x})=S_{1}(\vec{x})\pm iS_{2}(\vec{x})$. 
It is easy to check by using the relations (\ref{19})
that $S_{\pm}^{\prime}(\vec{x})$, 
$S_{3}^{\prime}(\vec{x})$ satisfy (\ref{22}). 
They also satisfy (\ref{21}) as seen using (\ref{18}).

The algebraic maps from the current algebras of ${\cal H}(4)$ to 
${\cal H}_{q}(4)$ can be obtained from the related manifolds (\ref{24}) 
and (\ref{27}). For instance,
\begin{equation}\label{35}
\ba{l}
S_{+}^{\prime}(\vec{x})=
\displaystyle\frac{\sinh\gamma(S_{3}(\vec{x})+\frac{1}{2})}
{(S_{3}(\vec{x})+\frac{1}{2})\gamma}S_{+}(\vec{x})\,,\\[4mm]
S_{-}^{\prime}(\vec{x})=\displaystyle\frac{\cosh\gamma(S_{3}(\vec{x})-
\frac{1}{2})}{\cosh\gamma}S_{-}(\vec{x})\,,
~~~S_{3}^{\prime}(\vec{x})=S_{3}(\vec{x})\,,
\ea
\end{equation}
where $S_{\pm}(\vec{x})$ and $S_{3}(\vec{x})$ are the generators of
the current algebra ${\cal H}(4)$ satisfying the relations (\ref{25}).
$S_{\pm}^{\prime}(\vec{x})$ and $S_{3}^{\prime}(\vec{x})$ are the 
generators of the current algebra ${\cal H}_{q}(4)$ satisfying the relations 
(\ref{28}). We also see that $S_{\pm}(\vec{x})$, $S_{3}(\vec{x})$ 
satisfy (\ref{24}) and $S_{\pm}^{\prime}(\vec{x})$, 
$S_{3}^{\prime}(\vec{x})$ satisfy (\ref{27}). 

Similarly, the current algebraic maps among other current extended 
algebras such as $SU(1,1)$ and $SU_{q}(1,1)$ can be studied by 
investigating their related current manifolds. 

\section{Geometric quantization of current algebras}

The first step towards geometric quantization is prequantization 
(cfr. \cite{wood,geom}). In our case the aim of prequantization
is to construct a linear monomorphism from the Poisson
algebra of $(M_N,\omega)$ to the space of linear operators on an appropriate
space by constructing the prequantization line bundle $L$. 
The system is still classical at this prequantization level.

In the process of quantization, one has to find a suitable classical 
counterpart
of the notion of a complete set of commuting observables. A natural choice 
is a set of $l=\displaystyle\frac{1}{2}dim (M)$ 
functions $f_{1},f_{2},...,f_{l}$ on $M$ such that
their Hamiltonian vector fields $X_{f_{1}},X_{f_{2}},...,X_{f_{l}}$ are 
complete. If one drops the assumption that the $f_{i}$ be real and globally
defined, one is led to the concept of a ``polarization'' $P$ 
of $(M_N,\omega)$. The complex linear combinations of
Hamiltonian vector fields $X_{f_{i}}$, $i=1,2,...,l$, span 
an involutive distribution $P$ on M such that
$dim(P)=\displaystyle\frac{1}{2}dim(M)$
and $\omega$ restricted to $P$ vanishes identically,
$\omega|_{P\times P}=0$ \cite{kirillov}.

A complex distribution $P$ satisfying above equations is called
a complex  Lagrangian distribution on $(M_N,\omega)$. A polarization of a 
symplectic manifold $(M_N,\omega)$ is a complex involutive Lagrangian 
distribution $P$ on $M_N$ such that $dim\{P\cap\bar{P}\}$ is constant, where 
$\bar{P}$ denotes the complex conjugate of $P$.

Let $\{X_i\}$, $i=1,...,\frac{1}{2}dim(M)$ be the basis of the 
polarization $P$. If the Hamiltonian vector
field of the observable $f$ on $M_N$ satisfies the relation
\be\label{36}
[X_{f},X_{i}](x)=\sum^{l}_{j=1}a^i\,_j (x) X_{j}(x)\,,~~~x\in M_N\,,
~~\forall i\,,
\ee
where $a^i\,_j (x)$ are smooth functions of $x$,
$f$ is said to be polarization preserving.
The quantum operator $\hat{f}$ associated to $f$ is then given 
by \cite{wood,geom}
\be\label{37}
\hat{f}=-i\hbar (X_{f}-\frac{i}{\hbar}\theta(X_{f}))+f
-\frac{1}{2}\hbar\sum_{j=1}^l a^j~_j\,,
\ee
where $\hbar=h/2\pi$ and $h$ is the Planck constant.
The quantum Hilbert space is the subspace of the section space of the
prequantization line bundle $L$ that is covariantly constant along the 
polarization.

In the following we take the current extended algebra of $SU(2)$ as an 
example.
As is shown in section 4 this algebra is related to the current 2-D sphere 
defined by equation (\ref{18}). Let  $S^{2}$ denote the latter manifold.
We set up a complex structure on $S^{2}$  by introducing two open sets
\begin{equation}\label{38}
U_{\pm}=\{z \in S^{2}\mid S_{0} \pm S_3(z)\not= 0\}  
\end{equation}
and two complex functions $ z_+(\vec{x})$ and $ z_-(\vec{x})$, 
$\vec{x}\in N$, on $ U_+ $ and $ U_-$ respectively,
\begin{equation}\label{39}
z_{\pm}(\vec{x})=\frac{S_1(\vec{x})\mp i S_2(\vec{x})}
{S_{0}\pm S_3(\vec{x})}\,.
\end{equation}
In $U_+\bigcap U_-$ we have
\begin{equation}\label{40}
z_{+}(\vec{x}) z_{-}(\vec{x})=1\,.
\end{equation}

From (\ref{39}) we get the expressions of $S_i(\vec{x})$, $i=1,2,3$, 
in terms of $z_+(\vec{x})$ and $z_-(\vec{x})$,
\begin{equation}\label{41}
\begin{array}{rcl}
S_1(\vec{x})&=&S_{0}\displaystyle\frac{z_{\pm}(\vec{x})
+\overline{z}_{\pm}(\vec{x})}
{1+z_{\pm}(\vec{x})\overline{z}_{\pm}(\vec{x})}\,, \\[5mm]
S_2(\vec{x})&=&\pm i S_{0}\displaystyle\frac{z_{\pm}(\vec{x})
-\overline{z}_{\pm}(\vec{x})}
{1+z_{\pm}(\vec{x})\overline{z}_{\pm}(\vec{x})}\,,\\[5mm]
S_3(\vec{x})&=&\pm S_{0}\displaystyle\frac{1-
z_{\pm}(\vec{x})\overline{z}_{\pm}(\vec{x})}{1+z_{\pm}(\vec{x})
\overline{z}_{\pm}(\vec{x})}\,.
\end{array}
\end{equation}

The symplectic form (\ref{20}) becomes
\begin{equation}\label{42}
\omega\mid U_{\pm}=\int_N\frac{-2 i S_{0}}
{\left(1+z_{\pm}(\vec{x})\overline{z}_{\pm}(\vec{x})\right)^2}\delta 
\overline{z}_{\pm}(\vec{x})
\wedge \delta z_{\pm}(\vec{x})d\mu(\vec{x})\,.
\end{equation}
Locally $\omega\mid U_{\pm}=\delta\theta_\pm$. From (\ref{42})
the symplectic one form $\theta_{\pm}$ reads
\begin{equation}\label{43}
\theta(\vec{x})\mid U_{\pm}=\int_N\frac{-2 i S_{0}}
{1+z_{\pm}(\vec{x})\overline{z}_{\pm}(\vec{x})} 
\overline{z}_{\pm}(\vec{x})\delta z_{\pm}(\vec{x})d\mu(\vec{x})\,.
\end{equation}

The Hamiltonian vector fields of $S_i(\vec{x})$ now take the form
\begin{equation}\label{44}
\begin{array}{rcl}
X_{S_1(\vec{x})}\mid U_{\pm}&=&-\displaystyle\frac{i}{2}
\left[\left(z_{\pm}^{2}(\vec{x})-1\right)\displaystyle\frac{\delta}
{\delta z_{\pm}(\vec{x})}+\left(1-\overline{z}_{\pm}^{2}(\vec{x})\right)
\displaystyle\frac{\delta}{\delta \overline{z}_{\pm}(\vec{x})}\right]\,,
\\[4mm]
X_{S_2(\vec{x})}\mid U_{\pm}&=&\pm\displaystyle\frac{1}{2}
\left[\left(z_{\pm}^{2}(\vec{x})+1\right)\displaystyle\frac{\delta}
{\delta z_{\pm}(\vec{x})}+\left(1+\overline{z}_{\pm}^{2}(\vec{x})\right)
\displaystyle\frac{\delta}{\delta \overline{z}_{\pm}(\vec{x})}\right]\,,
\\[4mm]
X_{S_3(\vec{x})}\mid U_{\pm}&=&\pm i \left(\overline{z}_{\pm}(\vec{x})
\displaystyle\frac{\delta}
{\delta \overline{z}_{\pm}(\vec{x})}-z_{\pm}(\vec{x})
\displaystyle\frac{\delta}{\delta z_{\pm}(\vec{x})}\right)\,.
\end{array}
\end{equation}

We suppose that the manifold $N$ has finite volume $V$.
The prequantization line bundle L exists if and only if  $V^{-1}h^{-1}\omega$
defines an integral de Rham cohomology class, i.e. the de Rham cohomology
class $[-V^{-1}h^{-1}\omega]$ of $-V^{-1}h^{-1}\omega$ should be
integrable \cite{geom}. This leads to the relation
$$ 
-V^{-1}h^{-1}\int_{S^{2}}\omega=2j ,\hspace{3mm} 2j\in\Nb.
$$
Hence we have 
\begin{equation}\label{45}
S_{0}=j \hbar\,,
\end{equation}
In the following discussions we shall, for simplicity, set $\hbar=1$

We take, as a suitable polarization, the linear 
frame fields of $z_{\pm}(\vec{x})$,
$$ 
X_{z_{\pm}(\vec{x})}= (2iS_0)^{-1}\left(1+z_{\pm}(\vec{x})
\overline{z}_{\pm}(\vec{x})\right)^2
\frac{\delta}{\delta \overline{z}_{\pm}(\vec{x})}.
$$
On $U_{+}\cap U_{-}$, $z_{+}(\vec{x})\ne 0$, we have
$X_{z_{-}(\vec{x})}=-z_{+}^{-2}(\vec{x})X_{z_{+}(\vec{x})}$.
Hence, $X_{z_+(\vec{x})}$ and  $X_{z_-(\vec{x})}$ span a complex 
distribution $P$ 
on $S^{2}$ and $P$ is a polarization of the symplectic manifold 
($S^{2}$,$\omega$). Moreover,
$$ 
 i \omega(X_{z_{\pm}(\vec{x})},\overline{X}_{z_{\pm}(\vec{x})})
 =\frac{1}{2S_{0}}(1+z_{\pm}(\vec{x})\overline{z}_{\pm}(\vec{x}))^2 >0\,.
$$
This means that (adapting a definition from \cite{geom})
$P$ is a complete strongly admissible positive polarization 
of ($S^{2}$, $\omega$) (in the sense of generalized functions).
       
From eqs.(\ref{37}), (\ref{41}) and (\ref{44}) we obtain (following the
procedures of geometric quantization) the following definition of the
corresponding quantum operators:
\begin{equation}\label{46}
\begin{array}{rcl}
\hat{S}_{1}(\vec{x})\mid U_{\pm}&
\equiv&-\displaystyle\frac{1}{2}(z_{\pm}^2(\vec{x})-1)
\displaystyle\frac{\delta}{\delta z_{\pm}(\vec{x})}
+j z_{\pm}(\vec{x})\,,\\[4mm]
\hat{S}_{2}(\vec{x})\mid U_{\pm}&
\equiv&\mp\displaystyle\frac{i}{2}(z_{\pm}^2(\vec{x})+1)
\displaystyle\frac{\delta}{\delta z_{\pm}(\vec{x})}
\pm ij z_{\pm}(\vec{x})\,,\\[4mm]
\hat{S}_{3}(\vec{x})\mid U_{\pm}&
\equiv&\pm(-z_{\pm}(\vec{x})\displaystyle
\frac{\delta}{\delta z_{\pm}(\vec{x})}+j)\,,
\end{array}
\end{equation}     
as operators acting on a suitable dense domain in the corresponding
quantum Hilbert space ${\cal H}$.
They give rise to the quantum current extended algebra of $SU(2)$, 
\begin{equation}\label{47}
\begin{array}{rcl}
[\hat{S}_{+}(\vec{x}),\hat{S}_{-}(\vec{y})]&=&
2\hat{S}_{3}(\vec{x})\delta(\vec{x}-\vec{y})\,,\\[4mm]
[\hat{S}_{3}(\vec{x}),\hat{S}_{\pm}(\vec{y})]&=&\pm\hat{ S}_{\pm}(\vec{x})
\delta(\vec{x}-\vec{y})\,,
\end{array}
\end{equation}
where $\hat{S}_{\pm}(\vec{x})=\hat{S}_{1}(\vec{x})\pm i\hat{S}_{2}(\vec{x})$,
$[A,B]$ denoting $AB-BA$. These relations have to be understood 
in the sense
of operator-valued generalized functions, the operators acting on 
a suitable dense domain in  ${\cal H}$.

From formula (\ref{46}) we directly have the following relation between the 
quantum operators $\hat{S}_{+}$, $\hat{S}_{-}$ and $\hat{S}_{3}$,
\be\label{48}
(j+\frac{1}{2})^2=\hat{S}_{+}(\vec{x})\hat{S}_{-}(\vec{x})+
(\hat{S}_{3}(\vec{x})-\frac{1}{2})^2\,,
\ee
This relation gives a ``quantum mechanical" analogue of the sphere $S^{2}$.

The geometric quantization of the current algebra of $SU_q(2)$ can be
studied in a similar way. Let $S_q^2$ denote the manifold defined by 
eq.(\ref{21}). 
Then the prequantization line bundle on $(S_q^2,\omega)$ exists iff when
$ -V^{-1}h^{-1}\int_{S_{q}^{2}}\omega=2j ,\hspace{3mm} 2j\in\Nb$,
where $\omega$ is given by formula 
(\ref{23}). This leads to an $S_\gamma$ in (\ref{21}) which takes the form
(for $\gamma\ne 0$)
$$
S_\gamma=\frac{\sinh\gamma S_0}{\sqrt{\gamma \sinh\gamma}}
=\frac{\sinh\gamma j}{\sqrt{\gamma \sinh\gamma}}\,.
$$
By geometric quantization we have the quantum operators of the
current extended algebra of $SU_q(2)$:
\begin{equation}\label{49}
\begin{array}{rcl}
\hat{S}_{1}(\vec{x})\mid V_{\pm}&=&\displaystyle\frac{1}{\sqrt{\gamma\sinh\gamma}}
(A_{\pm}+B_{\pm})\,,~~~~
\hat{S}_{2}(\vec{x})\mid V_{\pm}=\displaystyle\frac{1}{\sqrt{\gamma\sinh\gamma}}
(A_{\pm}-B_{\pm})\,,\\[4mm]
\hat{S}_{3}(\vec{x})\mid V_{\pm}&=&\pm\left(-z_{\pm}
(\vec{x})\displaystyle\frac{\delta}{\delta z_{\pm}(\vec{x})}
+j\right)\,,
\end{array}
\end{equation}     
where
$$
A_{\pm}\equiv\cosh\left(\displaystyle\frac{\gamma}{2}z_{\pm}(\vec{x})
\displaystyle\frac{\delta}{\delta z_{\pm}(\vec{x})}\right)
z_{\pm}(\vec{x})\sinh\left(\displaystyle\frac{\gamma}{2}(-z_{\pm}(\vec{x})
\displaystyle\frac{\delta}{\delta z_{\pm}(\vec{x})}+2j)\right)\,,
$$
$$
B_{\pm}\equiv\cosh\left(\displaystyle\frac{\gamma}{2}(-z_{\pm}(\vec{x})
\displaystyle\frac{\delta}{\delta z_{\pm}(\vec{x})}
+2j)\right)\displaystyle\frac{1}{z_{\pm}(\vec{x})}
\sinh\left(\displaystyle\frac{\gamma}{2}
z_{\pm}(\vec{x})\displaystyle\frac{\delta}{\delta z_{\pm}(\vec{x})}\right)
$$
(in the sense of operator-valued generalized functions on a dense
domain in  ${\cal H}$),
where the open sets $V_{\pm}$ on $S_q^2$ is defined by
$$
V_{\pm}=\left\{z \in \left. S^{2}_q\right\vert S_{\gamma} \pm 
\frac{\sinh\gamma S_3(z)}{\sqrt{\gamma \sinh\gamma}}\not= 0\right\}\,.  
$$

They satisfy (in the same sense as above) 
the commutation relations of the current extended 
quantum algebra of $SU_{q}(2)$,
\begin{equation}\label{50}
\begin{array}{rcl}
[\hat{S}_{+}(\vec{x}),\hat{S}_{-}(\vec{y})]&=&
\displaystyle\frac{\sinh(\gamma)}{\gamma} 
\displaystyle\frac{\sinh 2\gamma \hat{S}_{3}(\vec{x})}{\sinh\gamma}
\delta(\vec{x}-\vec{y})\,,\\[4mm]
[\hat{S}_{3}(\vec{x}),\hat{S}_{\pm}(\vec{y})]&=&\pm\hat{S}_{\pm}(\vec{x}) 
\delta(\vec{x}-\vec{y})\,,
\end{array}
\end{equation}
where $\hat{S}_{\pm}(\vec{x})=\hat{S}_{1}(\vec{x})\pm i\hat{S}_{2}(\vec{x})$.
This quantum current algebra is isomorphic to the classical Poisson
current algebra (\ref{22}). One may redefine $\hat{S}_{\pm}(\vec{x})$
by multiplying it by a constant factor $\sqrt{\frac{\sinh(\gamma)}{\gamma}}$ 
so as to get the usual form of commutation relations.

We remark that without setting $\hbar$ to be unit, 
eq.(\ref{50}) takes the form
$$
\begin{array}{rcl}
[\hat{S}_{+}(\vec{x}),\hat{S}_{-}(\vec{y})]&=&
\displaystyle\frac{\sinh(\gamma\hbar)}{\gamma} 
\displaystyle\frac{\sinh 2\gamma \hat{S}_{3}(\vec{x})}{\sinh\gamma}
\delta(\vec{x}-\vec{y})\,,\\[4mm]
[\hat{S}_{3}(\vec{x}),\hat{S}_{\pm}(\vec{y})]&=&\pm\hbar\hat{S}_{\pm}
(\vec{x})\delta(\vec{x}-\vec{y})\,.
\end{array}
$$
From this it is manifest that the $\hbar$-quantization and the q-deformation 
are independent operations.

In terms of the expressions (\ref{49}) and (\ref{50}) the quantum operators 
satisfy the following equation,
\be\label{51}
\hat{S}_{+}(\vec{x})\hat{S}_{-}(\vec{x})
+\frac{\sinh^2\gamma(\hat{S}_{3}(\vec{x})-\frac{1}{2})}
{\gamma\sinh\gamma}
=\frac{\sinh^2\gamma(j+\frac{1}{2})}{\gamma\sinh\gamma}~,
\ee
This is the quantum version of the manifold (\ref{21}).

The quantization of other current algebras can be discussed in a similar way. 
In addition, from the quantum version of the related manifolds, the quantum
current algebraic maps can be obtained. For instance, let 
$\hat{S}_{i}(\vec{x})$ (resp. $\hat{S}_{i}^\prime(\vec{x})$) be the quantum
operators of quantum current algebra of $SU(2)$ (resp. $SU_q(2)$)
satisfying eq.(\ref{48}) (resp. (\ref{51})). We set
$\hat{S}_{3}^{\prime}(\vec{x})=\hat{S}_{3}(\vec{x})$. Then
eq.(\ref{51}) can be rewritten as
$$
\ba{rcl}
\hat{S}_{+}^\prime(\vec{x})\hat{S}_{-}^\prime(\vec{x})
&=&\displaystyle\frac{\sinh^2\gamma(j+\frac{1}{2})}{\gamma\sinh\gamma}
-\displaystyle\frac{\sinh^2\gamma(\hat{S}_{3}(\vec{x})-\frac{1}{2})}
{\gamma\sinh\gamma}\\[4mm]
&=&\displaystyle\frac{1}{\sqrt{\gamma\sinh\gamma}}\hat{S}_{+}(\vec{x})
\displaystyle\frac{\sinh\gamma(j-\hat{S}_{3}(\vec{x}))}
{j-\hat{S}_{3}(\vec{x})}
\displaystyle\frac{1}{\sqrt{\gamma\sinh\gamma}}\hat{S}_{-}(\vec{x})
\displaystyle\frac{\sinh\gamma(j+\hat{S}_{3}(\vec{x}))}
{j+\hat{S}_{3}(\vec{x})}~,
\ea
$$
where eq.(\ref{48}) has been used. Hence we have
\be\label{52}
\hat{S}_{\pm}^{\prime}(\vec{x})=
\displaystyle\frac{1}{\sqrt{\gamma\sinh\gamma}}\hat{S}_{\pm}(\vec{x})
\displaystyle\frac{\sinh\gamma(j\mp\hat{S}_{3}(\vec{x}))}
{j\mp\hat{S}_{3}(\vec{x})}\,,~~~~
\hat{S}_{3}^{\prime}(\vec{x})=\hat{S}_{3}(\vec{x})\,.
\ee
Therefore from the ``quantum" current manifold (\ref{48}) of the
current extended quantum algebra $SU(2)$ 
and the ``quantum" current manifold (\ref{51}) of the
current extended quantum algebra $SU_q(2)$,
we get the quantum algebraic maps from the current extended algebras
$SU(2)$ to $SU_q(2)$, which is formally the same as
the classical Poisson algebraic maps (\ref{34}). Here
$\hat{S}_{3,\pm}^{\prime}(\vec{x})$ satisfy the relations
(\ref{51}) while $\hat{S}_{3,\pm}(\vec{x})$ satisfy (\ref{48}), as
guaranteed by Theorem 3.

\section{Conclusion and Remarks}

We have shown that there is a one to one correspondence (up to
algebraic equivalence) between Poisson current algebras of
maps from a Riemannian manifold $N$ to Poisson algebras with 
three generators and 2-D current manifolds.
This gives rise to 
a general description of current algebraic structures on 2-D manifolds. 
A geometric meaning of q-deformation of current algebras emerges
in terms of the corresponding manifolds. Also the Poisson and quantum 
algebraic maps can easily be handled in terms of the associated 
2-D current manifolds. 

The Hopf algebraic structures of current extended algebras can be studied 
by the quantum operators of these algebras. Recall that mathematically a Hopf 
algebra $A$ is an algebra with multiplication $m~:~~A\otimes A\to A$,
comultiplication (coproduct) $\Delta~:~~A\to A\otimes A$,
antipode $\eta~:~~A\to A$ and counit $\epsilon~:~~A\to K$ for a given 
field $K$. $\Delta$ is an algebra homomorphism $\Delta(ab)=\Delta(a)\Delta(b)$
and $\eta$ is an algebra anti-homomorphism $\eta(ab)=\eta(b)\eta(a)$
for $a,b\in A$. For these concepts see e.g. \cite{qgh}.
Here we give the Hopf algebraic structures for the
current extended algebra of $SU_q(2)$:
\be\label{aa2}
\begin{array}{l}
\Delta(\hat{S}_{3}(\vec{x}))=\hat{S}_{3}(\vec{x})\otimes {\bf 1} 
+{\bf 1}  \otimes\hat{ S}_{3}(\vec{x}),\\[3mm]
\Delta(\hat{ S}_{\pm}(\vec{x}))=\hat{S}_{\pm}(\vec{x})\otimes 
\displaystyle e^{\gamma\hat{ S}_{3}(\vec{x})}+
\displaystyle e^{-\gamma\hat{ S}_{3}(\vec{x})}\otimes
\hat{ S}_{\pm}(\vec{x}),\\[3mm]
\varepsilon({\bf 1})=1,\hspace{4mm} \varepsilon(\hat{S}_{\pm}(\vec{x}))=
\varepsilon(\hat{S}_{3}(\vec{x}))=0,\\[3mm]
\eta(\hat{S}_{\pm}(\vec{x}))=-\displaystyle 
e^{\gamma\hat{ S}_{3}(\vec{x})}\hat{S}_{\pm}(\vec{x})
\displaystyle e^{-\gamma
\hat{ S}_{3}(\vec{x})},\hspace{4mm}   
\eta(\hat{S}_{3}(\vec{x}))=-\hat{S}_{3}(\vec{x})\,.
\end{array}
\ee
These operations conserve the quantum current algebraic relations 
(\ref{50}). For instance,
$$
\begin{array}{rcl}
[\Delta\hat{S}_{+}(\vec{x}),\Delta\hat{S}_{-}(\vec{y})]&=&
\displaystyle\frac{\sinh(\gamma)}{\gamma} 
\displaystyle\frac{\sinh 2\gamma \Delta\hat{S}_{3}(\vec{x})}{\sinh\gamma}
\delta(\vec{x}-\vec{y})\,,\\[4mm]
[\Delta\hat{S}_{3}(\vec{x}),\Delta\hat{S}_{\pm}(\vec{y})]&=&
\pm\Delta\hat{S}_{\pm}(\vec{x}) \delta(\vec{x}-\vec{y})\,,
\end{array}
$$
(in the sense of operator-valued generalized functions on a dense 
domain in the quantum Hilbert space ${\cal H}$).
When the deformation parameter $\gamma$ approaches zero, (\ref{aa2}) becomes
formally the Hopf algebraic structures of quantum current extended 
algebra of $SU(2)$ and the comultiplication becomes commutative.

In this paper we obtained the quantum operator expressions of the current 
extend\-ed algebras through geometric quantization. In particular we have 
constructed here the representations of the current extended quantum algebra
$SU_q(2)$, see (\ref{50}). Representations of the current extended
quantum algebras can also be obtained in the form of highest weight 
representations based on continuous tensor product representations\cite{af2}.
It would be interesting to study the relations between the latter 
representations and the representations obtained in this paper.

In addition, by generalizing the 2-D current manifolds to Grassmannian 
manifolds, one can discuss BRST structures on 2-D current manifolds 
in terms of symplectic geometry and geometric quantization, extending the 
work we have done before for 2-D manifolds \cite{af1}.

\vspace{2.5ex}
ACKNOWLEDGEMENTS: We would like to thank Dr. A. Daletskii for
stimulating discussions. We also thank the A.v. Humboldt
Foundation for the financial support given to the second named author.

\end{document}